\documentclass[a4paper]{article}

\author{G. Gubbiotti\footnote{e-mail:
gubbiotti@mat.uniroma3.it},$\quad$  C. Scimiterna\footnote{e-mail:
scimiterna@fis.uniroma3.it}, $\quad$ D. Levi\footnote{e-mail:
decio.levi@roma3.infn.it}
\\
 Dipartimento di Matematica e Fisica, Universit\`a degli Studi Roma Tre,\\ e Sezione INFN di Roma Tre,\\ Via della Vasca Navale 84, 00146 Roma (Italy)
 }
\title{The non autonomous YdKN equation and generalized 
symmetries of Boll equations.}
\usepackage{rotating,color}

\usepackage{cite, amssymb}
\usepackage{braket}
\usepackage{amsmath}
\usepackage{booktabs}
\usepackage{amscd}
\usepackage[Q=yes]{examplep}
\usepackage{multirow}
\usepackage{rotating}
\usepackage{array}
\usepackage{longtable}
\usepackage{pdflscape}
\usepackage{array}
\newcolumntype{C}{>{\displaystyle} c <{}}

\newcommand{\Hvier}{$H^{4}$}
\newcommand{\Hsechs}{$H^{6}$}

\newcommand{\half}{\frac{1}{2}}
\newcommand{\Z}{\mathbb{Z}}

\def\bea{\begin{eqnarray}}
\def\eea{\end{eqnarray}}

\newcommand{\diff}[2]{\frac{\mathrm{d} #1}{\mathrm{d} #2}}

\newcommand{\tHeq}[2][]{${}_{t}H^{\varepsilon #1}_{#2}$}

\newcommand{\hX}{\widehat{X}}
\newcommand{\hY}{\widehat{Y}}
\newcommand{\Fp}[1]{F^{(+)}_{#1}}
\newcommand{\Fm}[1]{F^{(-)}_{#1}}

\newcommand{\Fppp}{\Fp{n}\Fp{m}}
\newcommand{\Fpmm}{\Fp{n}\Fm{m}}
\newcommand{\Fmpm}{\Fm{n}\Fp{m}}
\newcommand{\Fmmp}{\Fm{n}\Fm{m}}

\newcommand{\unm}{u_{n,m}}
\newcommand{\unpm}{u_{n+1,m}}
\newcommand{\unmp}{u_{n,m+1}}

\newcommand{\unmm}{u_{n-1,m}}

\newcommand{\ud}{\mathrm{d}}

\newcounter{rmk}
\renewcommand{\thermk}{\arabic{rmk}}
{\refstepcounter{rmk}\vspace{6pt}\noindent\ignorespaces\textbf{Remark
\thermk:}}{\vspace{6pt}\par}

\renewcommand{\epsilon}{\varepsilon}

\allowdisplaybreaks

\begin{document}

\maketitle
\begin{abstract}
In this paper we study the integrability of a class of nonlinear non autonomous quad graph equations compatible around the cube introduced by Boll. We show that all these equations possess three point generalized symmetries which are subcases of either the Yamilov discretization of the Krichever--Novikov equation   or of its non autonomous extension. We also prove that all those symmetries are integrable as pass the algebraic entropy test. 
\end{abstract}
\section{Introduction}

In 1983 Ravil I. Yamilov \cite{Yamilov1983} classified all differential difference equations of the class $\dot u_n = f(u_{n-1}, u_n, u_{n+1}) $ using the generalized symmetry method. From the generalized symmetry method one obtains integrability conditions which allow  to check whether a given equation is integrable.
Moreover in many cases these conditions enable us to classify equations, i.e. to obtain complete
lists of integrable equations belonging to a certain class. As integrability conditions are only necessary conditions for the
existence of generalized symmetries and/or conservation laws, one then has to prove that the
equations of the resulting list really possess generalized symmetries and conservation laws of sufficiently high order.
One mainly construct them using Miura-type transformations and master symmetries, proving the existence of Lax pairs \cite{Yamilov1984,Yamilov2006}. 
The result of Yamilov classification, up to Miura transformation, is the Toda equation and the so called Yamilov discretization of the Krichever Novikov equation (YdKN), a differential difference equation depending on 6 arbitrary coefficients:
\begin{equation}
    \diff{q_{k}}{t} = \frac{A(q_{k})q_{k+1}q_{k-1}
    + B(q_{k})(q_{k+1}+q_{k-1}) + C(q_{k})}{q_{k+1}-q_{k-1}},
    \label{eq:ydkn}
\end{equation}
where:
\begin{subequations}
\begin{align}
    A(q_k) &= a q_{k}^{2} + 2 b q_{k} + c,
    \\
    B(q_k) &= b q_{k}^{2} + d q_{k} + e,
    \\
    C(q_k) &= c q_{k}^{2} + 2 e q_{k} + f.
\end{align}
\label{eq:abcydkn}
\end{subequations}
The integrability of (\ref{eq:ydkn}) is proven by the existence of point symmetries \cite{LeviWinternitzYamilov} and of a master symmetries \cite{Yamilov2006} from which one is able to construct an infinite hierarchy of generalized symmetries. The problem of finding the B\"acklund transformation and Lax pair in the general case seems to be still open. 

In \cite{LeviYamilov1997} the authors constructed a set of five conditions necessary for the existence of
generalized symmetries for a class of differential-difference equations depending
only on nearest neighbouring interaction. They used the conditions to propose the integrability of the following  non autonomous generalization of the YdKN:
\begin{equation}
    \diff{q_{k}}{t} = \frac{A_{k}(q_{k})q_{k+1}q_{k-1}
    + B_{k}(q_{k})(q_{k+1}+q_{k-1}) + C_{k}(q_{k})}{q_{k+1}-q_{k-1}},
    \label{eq:naydkn}
\end{equation}
where the now $k$-dependent coefficients are given by:
\begin{subequations}
\begin{align}
    A_{k}(q_k)&= a q_{k}^{2} + 2 b_{k} q_{k} + c_{k},
    \\
    B_{k}(q_k) &= b_{k+1} q_{k}^{2} + d q_{k} + e_{k+1},
    \\
    C_{k} (q_k)&= c_{k+1} q_{k}^{2} + 2 e_{k} q_{k} + f,
\end{align}
    \label{eq:akbkcknaydkn}
\end{subequations}
\noindent with $b_{k}$, $c_{k}$ and $e_{k}$ 2-periodic functions.  Eq. (\ref{eq:naydkn}) has conservation laws of second and third  order
 and  two generalized local symmetries
of order $i$ and $i+1$, with $i<4$.

It was proved in \cite{LeviPetreraScimiterna2008} that the three point symmetries
of the equations belonging to the so-called ABS classification \cite{ABS2003},
found systematically in \cite{Hydon2007}, are all  particular cases of
 \eqref{eq:ydkn}. Here in this note we will show that the
three point generalized symmetries of all the equations coming from the classification of Boll 
\cite{ABS2009,Boll2011,Boll2012a,Boll2012b}, which extends the ABS
one \cite{ABS2003}, are all particular cases of the YdKN or the non autonomous YdKN. In particular we will present  the symmetries of all the classes
of equations \Hvier~and \Hsechs, noting that the symmetries of the rhombic
\Hvier \, were found firstly in \cite{XenitidisPapageorgiou2009}. For the remaining classes
of equations namely the trapezoidal \Hvier~and the \Hsechs~equations,
which were found to be \emph{linearizable} in \cite{GubScimLev2015},
this is the first time that their generalized symmetries are presented.
Furthermore we also present a new suggestion for the integrability
of the non-autonomous YdKN \eqref{eq:naydkn} based on the algebraic entropy test
and use the same criterion to prove the integrability of the other  non autonomous equations  of the $H^4$ and $H^6$ classes.

In Section 2 we present the three point generalized symmetries of the  $H^4$ and $H^6$ classes and identify them with subcases of the YdKN or of its non autonomous extension. In Section 3 we compute the algebraic entropy for the non autonomous YdKN and its subcases obtained before while in Section 4 we present some brief conclusions.

\section{Three point generalized symmetries and their identification}

In this Section we consider the various classes of equation coming from the classification of Boll 
\cite{ABS2009,Boll2011,Boll2012a,Boll2012b} as presented in \cite{GubScimLev2015}, show their
symmetries and show the identification of the fluxes of such symmetries
with the YdKN and its non autonomous extension.

\subsection{Rhombic \Hvier~equations}

Once written on the $\Z^{2}_{(n,m)}$ lattice,  according to \cite{XenitidisPapageorgiou2009}, the three equations
belonging to this class have the form:
\begin{subequations}
    \begin{align}
        _{r}H_{1}^\epsilon &\colon
        \begin{aligned}[t]
            &\phantom{+}(u_{n,m}-u_{n+1,m+1})\, (u_{n+1,m}-u_{n,m+1})\, -\,(\alpha \,- \, \beta)\\
        &+\epsilon (\alpha - \beta)\left(\Fp{n+m} \,u_{n+1,m} u_{n,m+1} 
        + \Fm{n+m} \,u_{n,m} u_{n+1,m+1}\right)= 0,
        \end{aligned}
        \label{eq:H1ae} 
        \\
        _{r}H_2^\epsilon &\colon
        \begin{aligned}[t]
            &\phantom{+}(u_{n,m}-u_{n+1,m+1})(u_{n+1,m}-u_{n,m+1}) + \\
        &+(\beta-\alpha) (u_{n,m}+u_{n+1,m}+u_{n,m+1}+u_{n+1,m+1}) 
        - \alpha^2 + \beta^2  \\
        &-  \epsilon\, (\beta-\alpha)^3  -\epsilon \,(\beta-\alpha) \,\left( 2 \Fm{n+m} u_{n,m} 
        + 2 \Fp{n+m} u_{n+1,m}+\alpha+\beta\right) \cdot \\
        &\cdot \left(  2 \Fm{n+m} u_{n+1,m+1} + 2 \Fp{n+m} u_{n,m+1} +\alpha+\beta\right )\,=\,0, 
        \end{aligned}
        \label{eq:H2ae}
        \\
        _{r}H_3^\epsilon &\colon
        \begin{aligned}[t]
            &\phantom{+}\alpha (u_{n,m} u_{n+1,m}+u_{n,m+1} u_{n+1,m+1}) \\
        &- \beta (u_{n,m} u_{n,m+1}+u_{n+1,m} u_{n+1,m+1}) + (\alpha^2-\beta^2) \delta  \\
        &-\, \frac{\epsilon (\alpha^2-\beta^2)}{\alpha \beta}
        \left(\Fp{n+m} \,u_{n+1,m} u_{n,m+1} + \Fm{n+m} \,u_{n,m} u_{n+1,m+1}\right) = 0,
        \end{aligned}
        \label{eq:H3ae}
    \end{align}
    \label{eq:H4rhombic}
\end{subequations}
where 
\begin{equation}
    F^{\pm}_{k} = \frac{1\pm(-1)^{k}}{2}, \quad k \in \Z.
    \label{eq:fpmk}
\end{equation}
Their generalized symmetries \cite{XenitidisPapageorgiou2009} are given
by:
\begin{subequations}
    \begin{align}
        \hX_{n}^{_{r}H_{1}^{\varepsilon}}
        & = \frac{1 - \varepsilon \left(\Fp{n+m}\unpm\unmm+\Fm{n+m}\unm^{2} \right)}{\unpm-\unmm}\partial_{\unm},
        \label{eq:XnH1e}
        \\
        \hX_{m}^{_{r}H_{1}^{\varepsilon}}
        & = \frac{1 - \varepsilon \left(\Fp{n+m}u_{n,m+1}u_{n,m-1}+\Fm{n+m}\unm^{2} \right)}{u_{n,m+1}-u_{n,m-1}}\partial_{\unm},
        \label{eq:XmH1e}
        \\
        \hX_{n}^{_{r}H_{2}^{\varepsilon}} &
        \begin{aligned}[t]
        &=\left [\frac{\left( 1 - 4\varepsilon\alpha\Fm{n+m} \right)\left( \unpm+\unmm \right) 
        - 4\varepsilon\Fp{n+m}\unpm\unmm}{\unpm-\unmm}+\right.
        \\
        &\left. +\frac{2\alpha - 4\varepsilon\alpha^{2}
        -4\varepsilon\Fm{n+m}\unm^{2}+\left( 1 - 4\varepsilon\alpha\Fm{n+m} \right)\unm}{\unpm-\unmm}\right ]\partial_{\unm}
        \end{aligned}
        \label{eq:XnH2e}
        \\
        \hX_{m}^{_{r}H_{2}^{\varepsilon}} &
        \begin{aligned}[t]
            &=\left [ \frac{\left( 1 - 4\varepsilon\beta\Fm{n+m} \right)\left( u_{n,m+1}+u_{n,m-1} \right) 
            - 4\varepsilon\Fp{n+m}u_{n,m+1}u_{n,m-1}}{u_{n,m+1}-u_{n,m-1}}+\right.
        \\
        &\left.+\frac{2\beta - 4\varepsilon\beta^{2}
        -4\varepsilon\Fm{n+m}\unm^{2}+\left( 1 - 4\varepsilon\beta\Fm{n+m} \right)\unm}{u_{n,m+1}+u_{n,m-1}}\right]\partial_{\unm}
        \end{aligned}
        \label{eq:XmH2e}
        \\
        \hX_{n}^{_{r}H_{3}^{\varepsilon}} &
        \begin{aligned}[t]
        &=\left[ \half\frac{\unm\left( \unpm+\unmm \right)  +2\delta\alpha}{\unpm-\unmm}-\right.
        \\
        &\left.-\frac{\varepsilon}{\alpha}\frac{\left( \Fp{n+m}\unpm\unmm + \Fm{n+m}\unm^{2} \right) }{\unpm-\unmm}\right]\partial_{\unm}
        \end{aligned}
        \label{eq:XnH3e}
        \\
        \hX_{m}^{_{r}H_{3}^{\varepsilon}} &
        \begin{aligned}[t]
            &=\left[ \half\frac{\unm\left( u_{n,m+1}+u_{n,m-1}\right)  +2\delta\beta}{u_{n,m+1}+u_{n,m-1}}-\right.
        \\
        &\left.-\frac{\varepsilon}{\beta}\frac{\left( \Fp{n+m}u_{n,m+1}u_{n,m-1} + \Fm{n+m}\unm^{2} \right) }{u_{n,m+1}+u_{n,m-1}}\right]\partial_{\unm}
        \end{aligned}
        \label{eq:XmH3e}
    \end{align}
    \label{eq:symmH4rhombic}
\end{subequations}

As stated in \cite{XenitidisPapageorgiou2009} the fluxes of the symmetries 
\eqref{eq:symmH4rhombic} are readily
identified with the corresponding cases of the non autonomous
YdKN equation \eqref{eq:naydkn}. Such identification is in this paper made
explicit by showing the appropriate values of the coefficients
of  \eqref{eq:naydkn} in Table \ref{tab:rhombicH4}.

\begin{table}
    \centering
    \[
        \begin{array}{CCCCCCCC}
        \toprule
        \text{Eq.} & k & a & b_{k} & c_{k} & d & e_{k} & f
        \\
        \midrule
        \multirow{2}{*}{$_{r}H_{1}^{\varepsilon}$} & n &
        0 & 0 & -\varepsilon \Fp{n+m} & 0 & 0 & 1
        \\
        & m &
        0 & 0 & -\varepsilon \Fp{n+m} & 0 & 0 & 1
        \\
        \multirow{2}{*}{$_{r}H_{2}^{\varepsilon}$} & n &
        0 & 0 & -4 \varepsilon \Fp{n+m} & 0 &
        1-4 \varepsilon \alpha \Fm{n+m} & 2 \alpha - 4 \varepsilon \alpha^{2}
        \\
        & m & 
        0 & 0 & -4 \varepsilon \Fp{n+m} & 0 &
        1-4 \varepsilon \beta\Fm{n+m} & 2 \beta - 4 \varepsilon \beta^{2}
        \\
        \multirow{2}{*}{$_{r}H_{3}^{\varepsilon}$} & n &
        0 & 0 & -\displaystyle\frac{\varepsilon \Fp{n+m}}{\alpha} & \half &
        0 & \delta \alpha
        \\
        & m & 
        0 & 0 & -\displaystyle\frac{\varepsilon \Fp{n+m}}{\beta} & \half &
        0 & \delta \beta
        \\
        \bottomrule
        \end{array}
    \]
    \caption{Identification of the coefficients in the symmetries of the rhombic
    \Hvier~equations with those of the non autonomous YdKN equation.}
    \label{tab:rhombicH4}
\end{table}

\subsection{Trapezoidal \Hvier~equations}

We now consider the trapezoidal \Hvier~equations,
which appeared in \cite{Boll2012a,Boll2012b} and whose
non-autonomous form was given in \cite{GubScimLev2015}:
\begin{subequations}
    \begin{align}
        _{t}H_{1}\colon &
        \begin{aligned}[t]
        &\left(u_{n,m}-u_{n+1,m}\right)  \left(u_{n,m+1}-u_{n+1,m+1}\right)-\\ 
        &-\alpha_{2}\epsilon^2\left({F}_{m}^{\left(+\right)}u_{n,m+1}u_{n+1,m+1}
        +{F}_{m}^{\left(-\right)}u_{n,m}u_{n+1,m}\right) -\alpha_{2}=0,
        \end{aligned}
        \label{eq:tH1e}
        \\
        _{t}H_{2}\colon &
        \begin{aligned}[t]
        &\left(u_{n,m}-u_{n+1,m}\right)\left(u_{n,m+1}-u_{n+1,m+1}\right)
        \\
        &+\alpha_{2}\left(u_{n,m}+u_{n+1,m}+u_{n,m+1}+u_{n+1,m+1}\right)
        \\
        &+\frac{\epsilon\alpha_{2}}{2} \left(2{F}_{m}^{\left(+\right)}u_{n,m+1}
        +2\alpha_{3}+\alpha_{2}\right)\left(2{F}_{m}^{\left(+\right)}u_{n+1,m+1}+2\alpha_{3}+\alpha_{2}\right)
        \\
        &+\frac{\epsilon\alpha_{2}}{2} \left(2{F}_{m}^{\left(-\right)}u_{n,m}+2\alpha_{3}
        +\alpha_{2}\right)\left(2{F}_{m}^{\left(-\right)}u_{n+1,m}+2\alpha_{3}+\alpha_{2}\right)
        \\
        &+\left(\alpha_{3}+\alpha_{2}\right)^2-\alpha_{3}^2-2\epsilon\alpha_{2}\alpha_{3}\left(\alpha_{3}+\alpha_{2}\right)=0
        \end{aligned}
        \label{eq:tH2e}
        \\
        _{t}H_{3}\colon &
        \begin{aligned}[t]
        &\alpha_{2}\left(u_{n,m}u_{n+1,m+1}+u_{n+1,m}u_{n,m+1}\right)
        \\
        &-\left(u_{n,m}u_{n,m+1}+u_{n+1,m}u_{n+1,m+1}\right)
        -\alpha_{3}\left(\alpha_{2}^{2}-1\right)\delta^2+
        \\
        &-\frac{\epsilon^2(\alpha_{2}^{2}-1)}{\alpha_{3}\alpha_{2}}
        \left({{F}_{m}^{\left(+\right)}u_{n,m+1}u_{n+1,m+1}
        +{F}_{m}^{\left(-\right)}u_{n,m}u_{n+1,m}}\right)=0,
        \end{aligned}
        \label{eq:tH3e}
    \end{align}
    \label{eq:trapezoidalH4}
\end{subequations}

We can easily calculate the three points
generalized symmetries of \tHeq{2} \eqref{eq:tH2e} and of
\tHeq{3} \eqref{eq:tH3e}:
\begin{subequations}
    \begin{align}
        \hX_{n}^{_{t}H_{2}^{\varepsilon}} &
        \begin{aligned}[t]
        &=\left[\frac{(\unm+\varepsilon\alpha_{2}^{2}\Fp{m})(u_{n+1,m}+u_{n-1,m}) 
       -u_{n+1,m}u_{n-1,m} }{%
        u_{n+1,m}-u_{n-1,m}}-\right.
        \\
        &\left.-\frac{u_{n,m}^{2}-2\epsilon{  \Fp{m}}\alpha_{2}^{2} u_{n,m}- 
        \alpha_{2}^{2}+4\epsilon{\Fp{m}}\alpha_{2}^{3}+8\epsilon\Fp{m}
        \alpha_{2}^{2}{\alpha_{3}}+\epsilon^{2}\Fp{m}\alpha_{2}^{4}
        }{u_{n+1,m}-u_{n-1,m}}\right]\partial_{\unm},
        \end{aligned}
        \label{eq:XnH2ep}
        \\
        \hX_{m}^{_{t}H_{2}^{\varepsilon}} &
        \begin{aligned}[t]
        &=\left[\frac{\left[\displaystyle\half - \varepsilon(\alpha_{2}+\alpha_{3})\Fp{m} \right]
        (u_{n,m+1}+u_{n,m-1})
        -\varepsilon \Fp{m}u_{n,m+1}u_{n,m-1} }{%
        u_{n,m+1}-u_{n,m-1}}-\right.
        \\
        &\left.-\frac{\varepsilon \Fm{m} u_{n,m}^{2}- 
        \left[1 - 2\varepsilon(\alpha_{2}+\alpha_{3}) \Fm{m}\right]  u_{n,m}+
        \alpha_{3} + \varepsilon \left( \alpha_{2}+\alpha_{3} \right)^{2}}{%
            u_{n,m+1}-u_{n,m-1}}\right]\partial_{\unm},
        \end{aligned}
        \label{eq:XmH2ep}
        \\
        \hX_{n}^{_{t}H_{3}^{\varepsilon}} &
        \begin{aligned}[t]
            &= \left[\frac{\displaystyle\half\alpha_{2}(1+\alpha_{2}^{2})\unm(u_{n+1,m}+u_{n-1,m}) 
        -\alpha_{2}^{2}u_{n+1,m}u_{n-1,m}}{%
        u_{n+1,m}-u_{n-1,m}}- \right.
        \\
        &\left.-\frac{\alpha_{2}^{2} \unm^{2}+\varepsilon^{2} \delta^{2} (1-\alpha^{2}_{2})^{2}\Fp{m}
        }{u_{n+1,m}-u_{n-1,m}}\right] \partial_{\unm},
        \end{aligned}
        \label{eq:XnH3ep}
        \\
        \hX_{m}^{_{t}H_{3}^{\varepsilon}} &
        \begin{aligned}[t]
            &=\left[\frac{\displaystyle\half\alpha_{3} \unm(u_{n,m+1}+u_{n,m-1})
        -\varepsilon^{2} \Fp{m}u_{n,m+1}u_{n,m-1} }{%
        u_{n,m+1}-u_{n,m-1}}-\right.
        \\
        &\left.-\frac{\varepsilon^{2}\Fm{m} \unm^{2} + \alpha_{3}^{2}\delta^{2}}{%
            u_{n,m+1}-u_{n,m-1}}\right]\partial_{\unm},
        \end{aligned}
        \label{eq:XmH3ep}
    \end{align}
    \label{eqn:symmetriestrapezoidalH4}
\end{subequations}
The symmetries in the $n$ and $m$ directions and the linearizations of the  \tHeq{1} equation \eqref{eq:tH1e} have been presented in \cite{Gallipoli15}. Their peculiarity is that they are determined by two arbitrary functions of one continuous variable and one lattice index and by arbitrary functions of the lattice indices. This is the first time that we find a lattice equation whose generalized symmetries depend on arbitrary functions. Almost surely this peculiarity is related to the very specific way in which  \tHeq{1} is linearizable. Here we  
present only the sub-cases which are related to the YdKN equation in its autonomous or non autonomous form.

The general symmetry in the $n$ direction is:

{\scriptsize\bea
&&\hat{X}_{n}^{{}_{t}H_{1}^\epsilon}=F_{m}^{(+)}\left\{\frac{\alpha_{2}\left(v^2+\epsilon^2\alpha_{2}^2\right)}{\left(r-v\right)\left(r+v\right)}B_{n}\left(\frac{\alpha_{2}}{r}\right)-\frac{\alpha_{2}\left(r^2+\epsilon^2\alpha_{2}^2\right)}{\left(r-v\right)\left(r+v\right)}B_{n-1}\left(\frac{\alpha_{2}}{v}\right)+\right.\label{Sator1}\\
&&\nonumber\left.+\left[u_{n,m}-\frac{\left(r^2+\epsilon^2\alpha_{2}^2\right)v}{\left(r-v\right)\left(r+v\right)}\right]\alpha+\gamma_{m}\right\}\partial_{u_{n,m}}+F_{m}^{(-)}\left[\frac{s^2t^2}{\left(s-t\right)\left(s+t\right)}\left(B_{n}\left(s\right)-\right.\right.\\
&&\nonumber\left.\left.-B_{n-1}\left(t\right)\right)-\frac{s^2t}{\left(s-t\right)\left(s+t\right)}\alpha+\delta_{m}\right]\left(1+\epsilon^2u_{n,m}^2\right)\partial_{u_{n,m}},\ \ \ r\doteq u_{n+1,m}-u_{n,m},\\
&&\nonumber s\doteq\frac{u_{n+1,m}-u_{n,m}}{1+\epsilon^2u_{n+1,m}u_{n,m}},\ \ \ t\doteq\frac{u_{n,m}-u_{n-1,m}}{1+\epsilon^2u_{n-1,m}u_{n,m}},\ \ \ \ \ \ v\doteq u_{n,m}-x_{u-1,m},
\eea}\\
\noindent where $B_{n}\left(x\right)$, $\gamma_{m}$ and $\delta_{m}$ are generic functions of their arguments and $\alpha$ is an arbitrary parameter. When $B_{n}\left(x\right)=-1/x$, $\alpha=\gamma_{m}=\delta_{m}=0$, we get
\bea \label{11}
\hat{X}_{n}^{{}_{t}H_{1}^\epsilon}=\left[\frac{\left(u_{n+1,m}-u_{n,m}\right)\left(u_{n,m}-u_{n-1,m}\right)}{u_{n+1,m}-u_{n-1,m}}-F_{m}^{(+)}\frac{\epsilon^2\alpha_{2}^2}{u_{n+1,m}-u_{n-1,m}}\right]\partial_{u_{n,m}}.
\eea

The general symmetry in the $m$ direction is:
\bea \label{h1}
      && \hX_{m}^{_{t}H_{1}^{\varepsilon}}  =  [\Fp{m}\, \left( 
        {B}_{m} \left( \frac {  u_{n,m+1}-u_{n,m-1}  }{1+{\epsilon}^{2}u_{{n,m+1}}u_{{n,m-1}}}
     \right)+{  \kappa}_{{m}} \right) 
     \\ \nonumber
     &&\qquad \qquad +{  \Fm{m}} \left( 1+
     {\epsilon}^{2}u_{{n,m}}^{2} \right) \left({C}_{m} \left(  u_{n,m+1}-u_{n,m-1}  \right) +{  \lambda}_{{m}} \right)]\partial_{u_{n,m}}.
 \eea
When $B_m(t)=1/t,C_m(t)=1/t$ and $\kappa_m=\lambda_m=0$ (\ref{h1}) becomes
\bea \label{ydkn}
\hX_{m}^{_{t}H_{1}^{\varepsilon}} 
        =[\Fp{m}\frac{1+\epsilon^2 u_{n,m+1}u_{n,m-1}}{u_{n,m+1}-u_{n,m-1}}+\Fm{m}\frac{1+\epsilon^2 u_{n,m}^2}{u_{n,m+1}-u_{n,m-1}}]\partial_{u_{n,m}}.
     \eea
Let us notice that the symmetries (\ref{eqn:symmetriestrapezoidalH4}, \ref{11})  in the $n$ 
direction are  sub-cases of the original YdKN equation. As $F^{(\pm)}_{m}$ 
depends on the other lattice index, it can be treated like a parameter
which is either $0$ or $1$. 

The explicit identification of the coefficients of the symmetries
\eqref{eqn:symmetriestrapezoidalH4}, (\ref{11}) and \eqref{ydkn} is shown in
Table \ref{tab:trapezodailH4}.

\begin{sidewaystable}
    \centering
    \[
        \begin{array}{CCCCCCCC}
        \toprule
        \text{Eq.} & k & a & b_{k} & c_{k} & d & e_{k} & f
        \\
        \midrule
        \multirow{2}{*}{\text{\tHeq[]{1}}} & n
        & 0 & 0 & -1 & 1 & 0 &   -\epsilon^2 \alpha_2^2\Fp{m}
        \\
       & m
        & 0 & 0 & \epsilon^2 \Fp{m} & 0 & 0 &  2
        \\
        \multirow{2}{*}{\text{\tHeq{2}}} & n
        & 0 & 0 & -1 & 1 & \varepsilon \alpha_{2}^{2} \Fp{m} &
        \alpha_{2}^{2} - 
        \varepsilon \alpha_{2}^{2}\left(4\alpha_{2} +8 \alpha_{3} 
        + \varepsilon \alpha_{2}^{2}  \right) \Fp{m}
        \\
        & m
        & 0 & 0 & -\varepsilon \Fp{m} & 0 & 
        \half - \varepsilon (\alpha_{2} + \alpha_{3}) \Fm{m} &
        -\alpha_{3} - \varepsilon\left( \alpha_{2}+\alpha_{3} \right)^{2}
        \\
        \multirow{2}{*}{\text{\tHeq{3}}} & n
        & 0 & 0 & -\alpha_{2}^{2} & \half\alpha_{2}(1+\alpha_{2}^{2})
        & 0 & - \varepsilon^{2} \delta^{2}\Fp{m} (1-\alpha_{2}^{2})^{2}
        \\
        & m & 0 & 0 & -\varepsilon^{2} \Fp{m} & \half\alpha_{3} & 0 & -\alpha_{3}^{2}\delta^{2}
        \\
        \bottomrule
        \end{array}
    \]
    \caption{Identification of the coefficients in the symmetries of the trapezoidal
    \Hvier~equations with those of the YdKN equation. In the direction $n$ the YdKN is autonomous while in the $m$ direction is non autonomous. Here the symmetries of  \tHeq[]{1} in the $m$ direction  are the subcase  (\ref{ydkn}) of (\ref{h1}) while those in the $n$ direction are the subcase (\ref{11}) of (\ref{Sator1}).}
    \label{tab:trapezodailH4}
\end{sidewaystable}

\subsection{\Hsechs~equations}

In this subsection we consider the equations of the family
\Hsechs~introduced in \cite{Boll2012a,Boll2012b}. We shall present
their non autonomous form on the lattice $\Z_{(n,m)}^{2}$ as given in \cite{GubScimLev2015}:
\begin{subequations}
    \begin{align}
        _{1}D_{2} &\colon
        \begin{aligned}[t]
        &\phantom{+}\left( F_{n+m}^{\left(-\right)}-\delta_{1} F_{n}^{\left(+\right)} F_{m}^{\left(-\right)}+\delta_{2} F_{n}^{\left(+\right)} F_{m}^{\left(+\right)}\right)u_{n,m}
        \\
        &+\left( F_{n+m}^{\left(+\right)}-\delta_{1} F_{n}^{\left(-\right)} F_{m}^{\left(-\right)}+\delta_{2} F_{n}^{\left(-\right)} F_{m}^{\left(+\right)}\right)u_{n+1,m}+
        \\ 
        &+\left( F_{n+m}^{\left(+\right)}-\delta_{1} F_{n}^{\left(+\right)} F_{m}^{\left(+\right)}+\delta_{2} F_{n}^{\left(+\right)} F_{m}^{\left(-\right)}\right)u_{n,m+1}
        \\
        &+\left( F_{n+m}^{\left(-\right)}-\delta_{1} F_{n}^{\left(-\right)} F_{m}^{\left(+\right)}+\delta_{2} F_{n}^{\left(-\right)} F_{m}^{\left(-\right)}\right)u_{n+1,m+1}+
        \\ 
        &+\delta_{1}\left( F_{m}^{\left(-\right)}u_{n,m}u_{n+1,m}+ F_{m}^{\left(+\right)}u_{n,m+1}u_{n+1,m+1}\right)
        \\
        &+ F_{n+m}^{\left(+\right)}u_{n,m}u_{n+1,m+1}
        + F_{n+m}^{\left(-\right)}u_{n+1,m}u_{n,m+1}=0,
        \end{aligned}
        \label{eq:1D2}
        \\
        _{2}D_{2} &\colon
        \begin{aligned}[t]
            &\phantom{+}\left(F_{m}^{\left(-\right)}-\delta_{1}F_{n}^{\left(+\right)}F_{m}^{\left(-\right)}+\delta_{2}F_{n}^{\left(+\right)}F_{m}^{\left(+\right)}-\delta_{1} \lambda F_{n}^{\left(-\right)}F_{m}^{\left(+\right)}\right)u_{n,m}
        \\
        &+\left(F_{m}^{\left(-\right)}-\delta_{1}F_{n}^{\left(-\right)}F_{m}^{\left(-\right)}+\delta_{2}F_{n}^{\left(-\right)}F_{m}^{\left(+\right)}-\delta_{1} \lambda F_{n}^{\left(+\right)}F_{m}^{\left(+\right)}\right)u_{n+1,m}
        \\
        &+\left(F_{m}^{\left(+\right)}-\delta_{1}F_{n}^{\left(+\right)}F_{m}^{\left(+\right)}+\delta_{2}F_{n}^{\left(+\right)}F_{m}^{\left(-\right)}-\delta_{1} \lambda F_{n}^{\left(-\right)}F_{m}^{\left(-\right)}\right)u_{n,m+1}
        \\
        &+\left(F_{m}^{\left(+\right)}-\delta_{1}F_{n}^{\left(-\right)}F_{m}^{\left(+\right)}+\delta_{2}F_{n}^{\left(-\right)}F_{m}^{\left(-\right)}-\delta_{1} \lambda F_{n}^{\left(+\right)}F_{m}^{\left(-\right)}\right)u_{n+1,m+1}
        \\
        &+\delta_{1}\left(F_{n+m}^{\left(-\right)}u_{n,m}u_{n+1,m+1}+F_{n+m}^{\left(+\right)}u_{n+1,m}u_{n,m+1}\right)
        \\ 
        &+F_{m}^{\left(+\right)}u_{n,m}u_{n+1,m}+F_{m}^{\left(-\right)}u_{n,m+1}u_{n+1,m+1}
        -\delta_{1}\delta_{2}\lambda=0,
        \end{aligned}
        \label{eq:2D2}
        \\
        _{3}D_{2} &\colon
        \begin{aligned}[t]
            &\phantom{+}\left(F_{m}^{\left(-\right)}-\delta_{1}F_{n}^{\left(-\right)}F_{m}^{\left(-\right)}+\delta_{2}F_{n}^{\left(+\right)}F_{m}^{\left(+\right)}-\delta_{1} \lambda F_{n}^{\left(-\right)}F_{m}^{\left(+\right)}\right)u_{n,m}
        \\
        &+\left(F_{m}^{\left(-\right)}-\delta_{1}F_{n}^{\left(+\right)}F_{m}^{\left(-\right)}+\delta_{2}F_{n}^{\left(-\right)}F_{m}^{\left(+\right)}-\delta_{1} \lambda F_{n}^{\left(+\right)}F_{m}^{\left(+\right)}\right)u_{n+1,m}
        \\
        &+\left(F_{m}^{\left(+\right)}-\delta_{1}F_{n}^{\left(-\right)}F_{m}^{\left(+\right)}+\delta_{2}F_{n}^{\left(+\right)}F_{m}^{\left(-\right)}-\delta_{1} \lambda F_{n}^{\left(-\right)}F_{m}^{\left(-\right)}\right)u_{n,m+1}
        \\
        &+\left(F_{m}^{\left(+\right)}-\delta_{1}F_{n}^{\left(+\right)}F_{m}^{\left(+\right)}+\delta_{2}F_{n}^{\left(-\right)}F_{m}^{\left(-\right)}-\delta_{1} \lambda F_{n}^{\left(+\right)}F_{m}^{\left(-\right)}\right)u_{n+1,m+1}
        \\
        &+\delta_{1}\left(F_{n}^{\left(-\right)}u_{n,m}u_{n,m+1}+F_{n}^{\left(+\right)}u_{n+1,m}u_{n+1,m+1}\right) 
        \\ 
        &+F_{m}^{\left(-\right)}u_{n,m+1}u_{n+1,m+1}
        +F_{m}^{\left(+\right)}u_{n,m}u_{n+1,m}-\delta_{1}\delta_{2}\lambda=0,
        \end{aligned}
        \label{eq:3D2}
        \\
        D_{3} &\colon
        \begin{aligned}[t]
            &\phantom{+}F_{n}^{\left(+\right)}F_{m}^{\left(+\right)}u_{n,m}+F_{n}^{\left(-\right)}F_{m}^{\left(+\right)}u_{n+1,m}
            +F_{n}^{\left(+\right)}F_{m}^{\left(-\right)}u_{n,m+1}
            \\
            &+F_{n}^{\left(-\right)}F_{m}^{\left(-\right)}u_{n+1,m+1}
            +F_{m}^{\left(-\right)}u_{n,m}u_{n+1,m}
            \\
            &+F_{n}^{\left(-\right)}u_{n,m}u_{n,m+1}+F_{n+m}^{\left(-\right)}u_{n,m}u_{n+1,m+1}+
        \\
        &+F_{n+m}^{\left(+\right)}u_{n+1,m}u_{n,m+1}+F_{n}^{\left(+\right)}u_{n+1,m}u_{n+1,m+1}
        \\
        &+F_{m}^{\left(+\right)}u_{n,m+1}u_{n+1,m+1}=0,
        \end{aligned}
        \label{eq:D3}
        \\
        _{1}D_{4} &\colon
        \begin{aligned}[t]
            &\phantom{+}\delta_{1}\left(F_{n}^{\left(-\right)}u_{n,m}u_{n,m+1}+F_{n}^{\left(+\right)}u_{n+1,m}u_{n+1,m+1}\right)+\\
            &+\delta_{2}\left(F_{m}^{\left(-\right)}u_{n,m}u_{n+1,m}+F_{m}^{\left(+\right)}u_{n,m+1}u_{n+1,m+1}\right)+\\
            &+u_{n,m}u_{n+1,m+1}+u_{n+1,m}u_{n,m+1}+\delta_{3}=0,
        \end{aligned}
        \label{eq:1D4}
        \\
        _{2}D_{4} &\colon
        \begin{aligned}[t]
            &\phantom{+}\delta_{1}\left(F_{n}^{\left(-\right)}u_{n,m}u_{n,m+1}+F_{n}^{\left(+\right)}u_{n+1,m}u_{n+1,m+1}\right)+
            \\
            &+\delta_{2}\left(F_{n+m}^{\left(-\right)}u_{n,m}u_{n+1,m+1}+F_{n+m}^{\left(+\right)}u_{n+1,m}u_{n,m+1}\right)+
            \\
            &+u_{n,m}u_{n+1,m}+u_{n,m+1}u_{n+1,m+1}+\delta_{3}=0.
        \end{aligned}
        \label{eq:2D4}
    \end{align}
    \label{eq:h6}
\end{subequations}

The three forms of the equation $D_{2}$ (\ref{eq:1D2},\ref{eq:2D2},\ref{eq:3D2}),
which we will collectively call $_{i}D_{2}$ assuming  $i$ 
in $\Set{1,2,3}$,
possess the following three points generalized symmetries in the $n$ direction and three points generalized symmetries in the $m$ direction:
\begin{subequations}
    \begin{align}
        \hX_{n}^{_{1}D_{2}} &
        \begin{aligned}[t]
            &=\left [\frac{\left(\Fpmm-\delta_{1}\Fppp\right)(\unmp+\unmm)}{%
                u_{n+1,m}-u_{n-1,m}}+\right.
            \\
            &+\frac{\left(\Fpmm-\delta_{1}\Fppp-\delta_{1}\delta_{2}\right)\Fmpm\unmm}{u_{n+1,m}-u_{n-1,m}}+
            \\
            &\left. +\frac{(\Fp{n+m}-\delta_{1}\Fp{m}-\delta_{1}\delta_{2}\Fppp)\unm+\delta_{2}\Fm{m}}{u_{n+1,m}-u_{n-1,m}}\right]\partial_{\unm},
        \end{aligned}
        \label{eq:Xn1D2}
        \\
        \hX_{m}^{{}_{1}D_{2}} &
        \begin{aligned}[t]
            &=\left[\frac{\delta_{1}\Fmpm u_{n,m+1}u_{n,m-1} + \Fm{n+m}(u_{n,m+1}+u_{n,m-1})}{u_{n,m+1}-u_{n,m-1}}+\right.
            \\
            &+\frac{\delta_{1}\Fp{m}u_{n,m+1} +\delta_{1}\delta_{2}\Fmpm u_{n,m-1}
            +\delta_{1}\Fmmp \unm^{2}}{u_{n,m+1}-u_{n,m-1}}+
            \\
            &+\frac{\left[\Fp{n+m} + \delta_{1}\left( \Fpmm-\Fmmp\right) 
            +\delta_{1}\delta_{2}\Fmmp\right]\unm }{u_{n,m+1}-u_{n,m-1}}-
            \\
            &\left.-\frac{\delta_{2}(\delta_{1}-1) \Fm{n} }{u_{n,m+1}-u_{n,m-1}}\right]\partial_{unm},
        \end{aligned}
        \label{eq:Xm1D2}
        \\
        \hX_{n}^{_{2}D_{2}} &
        \begin{aligned}[t]
            &=\left[\frac{\left(\Fmmp\delta_1+\Fmmp \delta_1\delta_2
            -\Fmmp\right)\unpm+\left( \Fppp\delta_1-\Fpmm \right)\unmm}{%
                u_{n+1,m}-u_{n-1,m}}+\right.
            \\
            &\left.+\frac{\left( \delta_{1}\Fm{n+m} - \Fm{m} 
            + \delta_{1}\delta_{2}\Fpmm \right)\unm-(\delta_{1}-1)\Fp{m}}{u_{n+1,m}-u_{n-1,m}}\right]\partial_{\unm},
        \end{aligned}
        \label{eq:Xn2D2}
        \\
        \hX_{m}^{{}_{2}D_{2}} &
        \begin{aligned}[t]
            &\left[\frac{\Fmmp\delta_{1} u_{n,m+1}u_{n,m-1}+
            \left( \delta_1\delta_2\Fmmp+\Fpmm \right) u_{n,m+1}
            }{u_{n,m+1}-u_{n,m-1}}+\right.
            \\
            &+\frac{+\left( \delta_1\Fppp+\Fmmp -\delta_1\Fmmp\right)u_{n,m-1}
            }{u_{n,m+1}-u_{n,m-1}}+
            \\
            &+\frac{\delta_{1}\Fmpm\unm^{2}
            +\left[ \Fpmm+(\delta_{2}-1)\Fmpm+\Fp{m}\right]\unm}{u_{n,m+1}-u_{n,m-1}}+
            \\
            &\left.+\frac{\delta_{2}(1-\delta_{2})\Fm{n} - \delta_{1}\lambda\Fp{n}}{u_{n,m+1}-u_{n,m-1}}\right]\partial_{\unm},
        \end{aligned}
        \label{eq:Xm2D2}
        \\
        \hX_{n}^{_{3}D_{2}} &
        \begin{aligned}[t]
            &=\left[\frac{\left(\delta_1\Fpmm+\delta_1\delta_2\Fpmm-\Fpmm \right)\unpm}{u_{n+1,m}-u_{n-1,m}}+\right.
            \\
            &+\frac{\left( \Fppp\delta_1-\Fmmp \right)\unmm}{u_{n+1,m}-u_{n-1,m}}+
            \\
            &\left.+\frac{\left( \delta_{1} \Fmmp\delta_2+\Fm{n}\delta_1-\Fm{m} \right)\unm
                +(1-\delta_1)\Fp{m}}{u_{n+1,m}-u_{n-1,m}}\right]\partial{\unm},
        \end{aligned}
        \label{eq:Xn3D2}
        \\
        \hX_{m}^{{}_{3}D_{2}} &
        \begin{aligned}[t]
            &=\left[\frac{\left( 1-\delta_1-\delta_1\delta_2 \right)\Fpmm u_{n,m+1}
            }{u_{n,m+1}-u_{n,m-1}}+\right.
            \\
            &+\frac{\left( \Fmmp-\Fppp\delta_1 \right) u_{n,m-1}+ \delta_{2}\Fm{n}}{u_{n,m+1}-u_{n,m-1}}+
            \\
            &+\frac{\left( \Fp{m}-\delta_{1}\Fp{n}-\delta_1\delta_2\Fppp\right)\unm
             }{u_{n,m+1}-u_{n,m-1}}-
                         \\
            &\left.-\frac{\lambda\delta_1(1-\delta_{1}-\delta_1\delta_2)\Fp{n}}{u_{n,m+1}-u_{n,m-1}} \right]\partial_{unm}.
        \end{aligned}
        \label{eq:Xm3D2}
    \end{align}
    \label{eq:symmiD2}
\end{subequations}
It can be readily proved that these symmetries are not non autonomous
YdKN equations \eqref{eq:naydkn}, however the equations $_{i}D_{2}$ possess
also the following point symmetries:
\begin{subequations}
    \begin{align}
        \hY_{1}^{_{1}D_{2}} &= \left( \Fppp+\Fpmm +\Fmpm \right) \unm\partial_{\unm},
        \label{eq:1D2p1}
        \\
        \hY_{2}^{_{1}D_{2}} &= \left[\delta_{1}\Fppp + [1-\delta_{1}(1 + \delta_{2})] \Fmpm
        +\Fpmm\right]\partial_{\unm},
        \\
        \hY_{1}^{_{2}D_{2}} &
        \begin{aligned}[t]
        &= \left[\left( \Fppp+\Fpmm +\Fmpm \right) \unm-\right.
        \\
        &\left. -\lambda\Fpmm + \lambda[1-\delta_{1}(1+\delta_{2})]\Fmmp\right]\partial_{\unm},
        \end{aligned}
        \label{eq:2D2p1}
        \\
        \hY_{2}^{_{2}D_{2}} &=\left[\delta_{1}\Fppp + \Fpmm
        [1-\delta_{1}(1+\delta_{2})]\Fmmp\right]\partial_{\unm},
        \label{eq:2D2p2}
        \\
        \hY_{1}^{_{3}D_{2}} &
        \begin{aligned}[t]
        &= \left[\left( \Fppp+\Fpmm +\Fmpm \right) \unm-\right.
        \\
        &\left.-\lambda\Fmmp + \lambda[1-\delta_{1}(1+\delta_{2})]\Fmmp\right]\partial_{\unm},
        \end{aligned}
        \label{eq:3D2p1}
        \\
        \hY_{2}^{_{3}D_{2}} &= \left[\delta_{1}\Fppp + [1-\delta_{1}(1+\delta_{2})]\Fpmm
        -\Fmmp\right]\partial_{\unm},
        \label{eq:3D2p2}
    \end{align}
    \label{eq:iD2point}
\end{subequations}

As the symmetries (\ref{eq:symmiD2}) are not in the
form of the YdKN equation \eqref{eq:naydkn}, we may look for 
a linear combination:
\begin{equation}
    \widehat{Z}_{j}^{_{i}D_{2}} = \hX_{j}^{_{i}D_{2}}
    + K_{1}\hY^{_{i}D_{2}}_{{1}} + K_{2} \hY^{_{i}D_{2}}_{{2}},
    \quad j =n,m;\;i=1,2,3
    \label{eq:combsymm}
\end{equation}
such that the resulting symmetries of equations $_{i}  D_{2}$ will be in the form \eqref{eq:naydkn}.
Indeed it turns out that this is the case and the resulting identification
with the proper constants $K_{1}$ and $K_{2}$ is displayed in Table \ref{tab:iD2}. The fact that the $_{i}D_{2}$ equations admit point symmetries and
generalized symmetries  makes them a unique case among
the equations of Boll classification.


\begin{sidewaystable}
    \centering
    \[
        \begin{array}{CCCCCCCCCC}
        \toprule
        \text{Eq.} & k & a & b_{k} & c_{k} & d & e_{k} & f & K_{1} & K_{2}
        \\
        \midrule
        \multirow{2}{*}{ \text{${}_{1}  D_{2}$}} & n
        & 0 & 0 & 0 & 0 & \displaystyle\half[\delta_{1}(1+\delta_{2})-1]\Fppp +\half\Fmpm-\half\Fmmp&
        -\delta_{2} \Fm{m}& 0 & -1/2
        \\
        & m
        & 0 & 0 & -\Fmpm\delta_{1} & 0 & \displaystyle
        \half(\delta_{1}(1-\delta_{2}-1)\Fmmp - \half\Fppp - \half\delta_{1}\Fmpm &
         \delta_{2}(\delta_{1}-1)\Fm{n} & 0 & -1/2
        \\
        \multirow{2}{*}{\text{$_{2}  D_{2}$}} & n
        & 0 & 0 & 0 & 0 & \half[1-\delta_{1}(1+\delta_{2})]\Fpmm+\half\Fmmp-\half\delta_{1}\Fmpm
        & (\delta_{1}-1) \Fp{m} & 0 & -1/2
        \\
        & m & 0 & 0 & -\delta_{1}\Fmmp & 0 
        & \half[\delta_{1}(1-\delta_{2})-1]\Fmpm -\half \Fppp -\half\delta_{1} \Fmpm
        & \delta_{2}\left[ \delta_{1}- 1\right]\Fm{n} + \lambda\delta_{1}\Fp{n}
        & 0 & -1/2
        \\
        \multirow{2}{*}{\text{$_{3}  D_{2}$}} & n
        & 0 & 0 & 0 & 0 & \half[\delta_{1}(1+\delta_{2})-1]\Fmmp+\half\Fpmm+\half\delta_{1}\Fmpm
        & (1-\delta_{1}) \Fp{m} & 0 & 1/2
        \\
        & m & 0 & 0 & 0 & 0 
        & \half[\delta_{1}(1-\delta_{2})-1]\Fppp -\half \Fpmm +\half\delta_{1} \Fmpm
        & \delta_{1}\lambda[-\delta_{1}(1+\delta_{2})]\Fp{n}-\delta_{2}\Fm{n}
        & 0 & 1/2
        \\
        \bottomrule
        \end{array}
    \]
    \caption{Identification of the coefficients of the symmetries
        of the $_{i}  D_{2}$ equations and value of the constants $K_{1}$
        and $K_{2}$ in \eqref{eq:combsymm} in order to obtain
         non autonomous YdKN equations.}
    \label{tab:iD2}
\end{sidewaystable}

The $D_{3}$ equation \eqref{eq:D3} admits only the following
three points generalized symmetries\footnote{Note that the equation $D_{3}$ \eqref{eq:D3}
is invariant under the exchange $n \leftrightarrow m$ so the symmetry $X_{m}^{D_{3}}$
\eqref{eq:XmD3} can be obtained from the symmetry $X_{n}^{D_{3}}$ \eqref{eq:XmD3} performing
such exchange.}:
\begin{subequations}
    \begin{align}
        \hX_{n}^{D_{3}} &
        \begin{aligned}[t]
        &= \left[\frac{\displaystyle\Fppp u_{n+1,m}u_{n-1,m}
        +\half\left( \Fm{m} - \Fmpm \right)\unm(u_{n+1,m}+u_{n-1,m})}{%
            u_{n+1,m}-u_{n-1,m}}+\right.
        \\
        &\left.+\frac{\Fmpm \unm^{2} + \left( \Fm{m} - \Fppp \ \right)\unm}{%
        u_{n+1,m}-u_{n-1,m}}\right]\partial_{\unm},
        \end{aligned}
        \label{eq:XnD3}
        \\
        \hX_{m}^{D_{3}} &
        \begin{aligned}[t]
        &= \left[\frac{\displaystyle\Fppp u_{n,m+1}u_{n,m-1}
        +\half\left( \Fm{n} - \Fpmm \right)\unm(u_{n,m+1}+u_{n,m-1})}{%
            u_{n,m+1}-u_{n,m-1}}+\right.
        \\
        &\left.+\frac{\Fpmm \unm^{2} + \left( \Fm{n} - \Fppp \ \right)\unm}{%
        u_{n,m+1}-u_{n,m-1}}\right]\partial_{\unm}
        \end{aligned}
        \label{eq:XmD3}
    \end{align}
and no point symmetries. Also the two forms of  $D_{4}$ possess only the following three point generalized symmetries:
    \begin{align}
        \hX_{n}^{_{1}D_{4}} &
        \begin{aligned}[t]
            &=\left[ \frac{\displaystyle-\delta_{1}\Fp{n} u_{n+1,m}u_{n-1,m}
        -\half\unm(u_{n+1,m}+u_{n-1,m})}{%
            u_{n+1,m}-u_{n-1,m}}+\right.
        \\
        &\left.+\frac{-\delta_{1}\Fm{n}\unm^{2} + \delta_{2}\delta_{3}\Fp{m}}{%
        u_{n+1,m}-u_{n-1,m}}\right]\partial_{\unm},
        \end{aligned}
        \label{eq:Xn1D4}
        \\
        \hX_{m}^{_{1}D_{4}} &
        \begin{aligned}[t]
            &= \left[\frac{\displaystyle \Fm{m} u_{n,m+1}u_{n,m-1}
        +\half\unm(u_{n,m+1}+u_{n,m-1})}{%
            u_{n,m+1}-u_{n,m-1}}+\right.
        \\
        &\left.+\frac{ \delta_{2}\Fp{m} \unm^{2} - \delta_{1}\delta_{3} \Fp{n}}{%
        u_{n,m+1}-u_{n,m-1}}\right]\partial_{\unm},
        \end{aligned}
        \label{eq:Xm1D4}
        \\
        \hX_{n}^{_{2}D_{4}} &
        \begin{aligned}[t]
            &=\left[ \frac{\displaystyle-\delta_{1}\delta_{2}\Fppp u_{n+1,m}u_{n-1,m}
        +\half\unm(u_{n+1,m}+u_{n-1,m})}{u_{n+1,m}-u_{n-1,m}}+\right.
        \\
        &\left.+\frac{-\delta_{1}\delta_{2}\Fmpm\unm^{2} + \delta_{3}}{%
        u_{n+1,m}-u_{n-1,m}}\right]\partial_{\unm},
        \end{aligned}
        \label{eq:Xn2D4}
        \\
        \hX_{m}^{_{2}D_{4}} &
        \begin{aligned}[t]
            &=\left[ \frac{\displaystyle \delta_{2}\Fp{n+m} u_{n,m+1}u_{n,m-1}
        +\half\unm(u_{n,m+1}+u_{n,m-1})}{u_{n,m+1}-u_{n,m-1}}+\right.
        \\
        &\left.+\frac{ \delta_{2}\Fm{n+m} \unm^{2} - \delta_{1}\delta_{3} \Fp{n}}{%
        u_{n,m+1}-u_{n,m-1}}\right]\partial_{\unm},
        \end{aligned}
        \label{eq:Xm2D4}
    \end{align}
    \label{eq:symmD3iD4}
\end{subequations}
and no point symmetries. Again the fluxes of the symmetries \eqref{eq:symmD3iD4}
can be readily identified with some specific form of the
non autonomous YdKN equations \eqref{eq:naydkn} and the explicit form
of the coefficients are shown in Table \ref{tab:D3iD4}.
\begin{sidewaystable}
    \centering
    \[
        \begin{array}{CCCCCCCC}
        \toprule
        \text{Eq.} & k & a & b_{k} & c_{k} & d & e_{k} & f
        \\
        \midrule
        \multirow{2}{*}{ \text{$D_{3}$}} & n
        & 0 & 0 & \Fppp & 0 & \half\left( \Fpmm+\Fmmp-\Fppp \right) & 0
        \\
        & m & 0 & 0 & \Fppp & 0 & \half\left( \Fmpm+\Fmmp-\Fppp \right) & 0
        \\
        \multirow{2}{*}{\text{$_{1}D_{4}$}} & n &
        0 & 0 & -\delta_{1}\left( \Fppp+\Fpmm \right) & -\half & 0 &
        \delta_{2}\delta_{3} \Fp{m}
        \\
        & m & 0 & 0 & \delta_{2}\left( \Fppp + \Fmpm \right) & \half &
        0 & -\delta_{1}\delta_{3} \Fp{n}
        \\
        \multirow{2}{*}{\text{$_{2}D_{4}$}} & n & 0 & 0 & -\Fppp\delta_{1}\delta_{2}
        & \half & 0 & \delta_{3}
        \\
        & m & 0 & 0 & \delta_{2}\left( \Fppp+\Fmmp \right) & \half & 0 & -\delta_{1}\delta_{3}\Fp{n} 
        \\
        \bottomrule
        \end{array}
    \]
    \caption{Identification of the coefficients of the symmetries \eqref{eq:symmD3iD4}
        for $D_{3}$, $_{1}D_{4}$ and $_{2}D_{4}$ with those of a non autonomous YdKN.}
    \label{tab:D3iD4}
\end{sidewaystable}

\section{Algebraic entropy for the non autonomous YdKN equation
    and its subcases.}

In the previous section we saw that the fluxes of all
the generalized three point
symmetries of the \Hvier \, and  \Hsechs \; equations
are eventually  related either to the YdKN (\ref{eq:ydkn}) or to
the non autonomous YdKN equation \eqref{eq:naydkn}.

It was remarked in the introduction that the non autonomous
YdKN equation \eqref{eq:naydkn} passes the necessary condition
for the integrability which is only an indication of the integrability of  such
class of equation. In this paper we have shown that they are symmetries of the 
\Hvier \, and  \Hsechs \; equations. Here we give
a further evidence that the non-autonomous YdKN might be an integrable
differential-difference equations based on the 
{algebraic entropy} test \cite{BellonViallet1999}.

We recall briefly how to compute the algebraic entropy in the
case of differential-difference equations of the form 
$\ud u_n / \ud t = f_{n}\left(u_{n+1},u_{n},u_{n-1} \right)$ 
\cite{DemskoyViallet2012,viallet2014}. 
First of all we assume that the equation is solvable for $u_{n+1}$ uniquely. 
This is a condition on $f_{n}$. Then,  starting from $n=0$, we compute  $u_{1}$ by substituting the
initial conditions:
\begin{equation}
    \Set{\frac{\ud^{k} u_{-1}}{\ud t^{k}},\frac{\ud^{k} u_{0}}{\ud t^{k}}}_{k=0}^{\infty}.
    \label{eq:ini}
\end{equation}
Knowing $u_{1}$ we can then calculate $u_{2}$ and so on.
We  define the degree of the iterate at the $l$-th step 
as the maximum between the degree of the numerator and of the numerator
of $u_{l}$ in the initial conditions \eqref{eq:ini}. A great simplification
in the explicit calculations is obtained if instead of a generic
initial condition one parametrizes the curve of the initial
condition rationally using the variable $t$:
\begin{equation}
    u_{-1} = \frac{A_{-1} t + B_{-1}}{A t + B}, \quad
    u_{0} = \frac{A_{0} t + B_{0}}{A t + B},
    \label{eq:inilin}
\end{equation}
and then compute the degrees in $t$.
Calculating $N$ iterates, for a sufficiently large positive integer
$N$, and constructing the generating function one can calculate the algebraic entropy
without calculating the entire sequence. For more details on how
the method is implemented see \cite{gubbiotti_thesis}.

We look for the sequence of degrees of the iterate map
for the non autonomous YdKN equation \eqref{eq:naydkn} and its particular cases  found
in the previous section.  We find for all the cases, except the symmetries
of  $_r H_{1}^{\varepsilon}$, the symmetries (\ref{11}, 
\ref{ydkn}) for \tHeq{1}  and the symmetries of the ${_{i}D_{2}}$ equations, the following values: 
\begin{equation}
    1, 1, 3, 7, 13, 21, 31, 43, 57, 73, 91, 111, 133, 157\dots
    \label{eq:naydkndegrees}
\end{equation}
This sequence has the following generating function:
\begin{equation}
    g(z) = \frac{1-2z+3z^{2}}{(1-z)^{2},}
    \label{eq:ydkngenfunc}
\end{equation}
which gives the following quadratic fit for the sequence \eqref{eq:naydkndegrees}:
\begin{equation}
    d_{l} = l(l-1)+1.
    \label{eq:naydkngrowthfit}
\end{equation}
For the symmetry in the $n$ direction \eqref{eq:H1ae} of the equation $_r H_{1}^{\varepsilon}$ we have the somehow different situation when the sequence growth is different according to the even or odd values of the $m$ variable:
\begin{subequations}
    \begin{align} \label{eq:20a}
m = 2k \quad &1,1,3,7,10,17,23,33,42,55,67,83,98,117\dots
 \\ \label{eq:20b}
m=2k+1 \quad &1,1,3,4,9,13,21,28,39,49,63,76,93,109 \dots  .
    \end{align}
    \label{eq:20}
\end{subequations}
These sequences have the following generating functions and asymptotic fits:
\begin{subequations}
    \begin{align}
        m = 2k, 
        \quad 
        &\begin{aligned}
        & g(z)=\frac{2 z^5 - 3 z^4 + 3 z^3 + z^2 - z + 1}{(1-z)^3(z + 1)},
        \\
        &d_l=\frac{3}{4}l^2 - l- \frac{5(-1)^l - 21}{8},    
        \end{aligned}
        \label{eq:21a}
        \\
        m=2k+1, 
        \quad 
        &\begin{aligned}
        & g(z)=\frac{(z^2 + z + 1)(2z^2 - 2z + 1)}{(1-z)^3(z + 1)},
        \\
        &d_l=\frac{3}{4}l^2 - \frac{3}{2}l- \frac{5(-1)^l - 19}{8}.
        \end{aligned}
        \label{eq:21b}
    \end{align}
    \label{eq:21}
\end{subequations}
The symmetry in the $m$ direction \eqref{eq:XmH1e} of the equation $_r H_{1}^{\varepsilon}$
has the same behaviour by exchanging $m$ with $n$ in formulae (\ref{eq:20}-\ref{eq:21}).
The $n$ directional symmetry   of the equation  \tHeq{1} has almost the same growth for $m$ odd as (\ref{eq:20b}, \ref{eq:21b}) however it is worthwhile to mention that the fit $d_l= \frac{3}{4} l^2 - \frac{5}{4}l + (-1)^l \frac{l}{4} + \frac{(-1)^l+15}{8}$ presents a term $l (-1)^l$, new in this kind of results. For $m$ even we have the same growth as (\ref{eq:naydkndegrees}). The $m$ directional symmetry   of the equation  \tHeq{1} has the same growth as the even one of \tHeq{1} (\ref{eq:20a}, \ref{eq:21a}).  For  the symmetries  (\ref{eq:combsymm}) of the ${_{i}  D_{2}}$ equations     we have  different growth according to the even or odd values of the $m$ or $n$ variables and  similar sequences or slightly lower than in the case of equations $H_{1}^{\varepsilon}$, however always corresponding to a quadratic asymptotic fit. 
 
 This shows that the whole family of the non autonomous YdKN
is integrable according to the algebraic entropy test.

For completeness let us  just mention that   the symmetries (\ref{eq:symmiD2}) of the $_{i}D_{2}$ equations have a  sequence growth of the same order  than those considered above, i.e. quadratic growth and thus null entropy.


\section{Conclusions}

In this note we constructed the symmetries of the equations belonging
to the Boll classification \cite{Boll2012a,Boll2012b} and showed that they are integrable (by the algebraic entropy test) and related to  particular
cases of the non autonomous YdKN equation \eqref{eq:naydkn} \cite{LeviYamilov1997}.
This was already known for the rhombic \Hvier~equations \cite{XenitidisPapageorgiou2009} and here we show  the explicit identification of the symmetries
obtained in that paper with the coefficients of the non autonomous YdKN equation.

We finally note that, as was proved in \cite{LeviPetreraScimiterna2008} for the YdKN (\ref{eq:ydkn}), 
no equation belonging to the Boll classification has a symmetry which corresponds to the general non autonomous YdKN equation (\ref{eq:naydkn}). In all the cases of the Boll classification one has $a=b_{k}=0$.

 In \cite{Xenitidis2009} it was shown that the $Q_V$ equation introduced by Viallet, possessing the Klein symmetry,
 \bea  \nonumber
    Q_{V}&:&{\it p_4}+{\it p_3}\, \left( u_{{n,m}}+{\it u}_{{n,m+1}}+u_{{n+1,m}}+{\it u}_{
{n+1,m+1}} \right) +\\ \nonumber &&+{\it p_{2,1}}\, \left( u_{{n,m}}u_{{n+1,m}}+{\it u}_{{n,m+1}}{
\it u}_{{n+1,m+1}} \right) +\\  \label{eq:TTX} && +{\it p_{2,2}}\, \left( u_{{n,m}}{\it u}_{{n,m+1}}+u_{{
n+1,m}}{\it u}_{{n+1,m+1}} \right) +\\ \nonumber && +{\it p_{2,0}}\, ( u_{{n,m}}{\it u}_{{n+
1,m+1}}+{\it u}_{{n,m+1}}u_{{n+1,m}} ) +\\ \nonumber &&+{\it p_1}\, ( u_{{n,m}}{\it u}
_{{n,m+1}}u_{{n+1,m}}+u_{{n,m}}u_{{n+1,m}}{\it u}_{{n+1,m+1}}+\\ \nonumber &&+u_{{n,m}}{\it u}_{{n,m+1}}
{\it u}_{{n+1,m+1}}+{\it u}_{{n,m+1}}u_{{n+1,m}}{\it u}_{{n+1,m+1}} ) +\\ \nonumber &&+{
\it p_0}\,u_{{n,m}}{\it u}_{{n,m+1}}u_{{n+1,m}}{\it u}_{{n+1,m+1}}
 = 0,  
\eea
admits a symmetry of the form of the YdKN:
\begin{equation}
    \hX_{n}^{\text{V}} = \frac {h}{u_{n+1,m} - u_{n-1,m}} 
    - \frac{1}{2} \partial_{u_{n+1,m}}h. 
    \label{eq:ttxsymm}
\end{equation}
where:
\bea \label{V}
h(u_{n,m}, u_{n+1,m}; {\it p_1, p_{2,i}, p_3, p_4}) &=& Q_{V}  \partial_{u_{n,m+1}} \partial_{u_{n+1,m+1}} Q_{V} +\\ \nonumber 
&& -\left(  \partial_{u_{n,m+1}} Q_{V} \right) \left(   \partial_{u_{n+1,m+1}} Q_{V}  \right).
\eea
The connection formulae between the coefficient of $Q_V$ and the YdKN \eqref{eq:ydkn} is:
\bea \label{QV-YdKN}
a&=&{\it p_1^2-p_{2,1} p_0}, \qquad b=\frac 1 2 [{\it p_1 (p_{2,0}+p_{2,2}-p_{2,1})-p_3 p_0}], \\ \nonumber c&=& {\it p_{2,0} p_{2,2}- p_3 p_1}, \qquad d=\frac 1 2 [{\it p_{2,2}^2-p_{2,1}^2+p_{2,0}^2 -p_0 p_4}], \\ \nonumber
e&=&\frac 1 2 [{\it p_3 (p_{2,2}-p_{2,1}+p_{2,0})-p_1 p_4}], \qquad f= p_3^2 - p_{2,1} p_4.
\eea
This is a set of coupled nonlinear algebraic equations between the 7 parameters ${\it p_i}$ of $Q_V$ and the 6 ones $(a,\cdots, f)$  of the YdKN. Eq. (\ref{QV-YdKN}) tells us that the YdKN with coefficients given by (\ref{QV-YdKN})  is a three point generalized symmetry of $Q_V$. If a solution of (\ref{QV-YdKN}) exists, i.e. one is able to express the ${\it p_i}$ in term of $(a,\cdots, f)$, then (a reparametrization of) $Q_V$ turns out to be a B\"acklund transformation of the YdKN \cite{Levi1981}.

\indent From the results obtained in this paper one is lead to conjecture a non autonomous generalization of the $Q_{V}$ equation. We have many possible ways of proposing such a  generalization. A first possibility is to generalize the original Klein symmetry:
\bea
&&Q\left(u_{n+1,m},u_{n,m},u_{n+1,m+1},u_{n,m+1};-(-1)^n,(-1)^m\right)=\label{KL} \\ &&\tau Q\left(u_{n,m},u_{n+1,m},u_{n,m+1},u_{n+1,m+1};(-1)^n,(-1)^m\right),\nonumber\\
&&Q\left(u_{n,m+1},u_{n+1,m+1},u_{n,m},u_{n,m+1};(-1)^n,-(-1)^m\right)=\nonumber \\ &&\tau^{\prime}Q\left(u_{n,m},u_{n+1,m},u_{n,m+1},u_{n+1,m+1};(-1)^n,(-1)^m\right),\nonumber
\eea
\noindent where $(\tau$, $\tau^{\prime})=\pm 1$ and $Q\left(x,u,y,z;(-1)^n,(-1)^m\right)$ is a multilinear function of its arguments with nonautonomous coefficients in the form of 2-periodic functions in $n$ and $m$, i. e. of the form $\alpha+\beta (-1)^n+\gamma (-1)^m+\delta (-1)^{n+m}$, with $\alpha$, $\beta$,  $\gamma$ and $\delta$   constants. This discrete symmetry is  shared by all of the Boll systems and in the autonomous case reduces to the the usual Klein symmetry.

 A second possibility  is  to ask the function $Q\left(x,u,y,z;(-1)^n,(-1)^m\right)$ to respect a strict Klein symmetry just as in  (\ref{eq:TTX}).   Choosing the coefficients for example as 
 $$p_{0}=1+(-1)^n, \; p_{1}=(-1)^n, \;p_{2,1}=-1+(-1)^n, \;p_{2,2}=(-1)^n,$$ $$p_{2,0}=1+2(-1)^n,\; p_{3}=1+(-1)^n, \; p_{4}=4+2(-1)^n,$$  (\ref{QV-YdKN}) provide a non autonomous YdKN. In this case, performing   the algebraic entropy test the equation turns out to be  integrable. Its generalized symmetries, however, are not necessarily in the form of a non autonomous YdKN equation. A different non autonomous choice of the coefficients of (\ref{eq:TTX}), such  that (\ref{QV-YdKN}) is satisfied for the coefficients of the non autonomous YdKN, gives, by the algebraic entropy test,  a non integrable  equation. 
 
  The proof of the existence of  a non autonomous generalization of  $Q_{V}$   together with the derivation of an effective B\"acklund transformation and Lax pair for the  YdKN and its non autonomous counterpart is work in progress.

\subsection*{Acknowledgments}
CS and DL  have been partly supported by the Italian Ministry of Education and Research, 2010 PRIN {\it Continuous and discrete nonlinear integrable evolutions: from water waves to symplectic maps}.

All the authors are supported   by INFN   IS-CSN4 {\it Mathematical Methods of Nonlinear Physics}.


\begin{thebibliography}{10}

\bibitem{ABS2003}
V.~E. Adler, A.~I. Bobenko, and Y.~B. Suris.
\newblock Classification of integrable equations on quad-graphs. the
  consistency approach.
\newblock {\em Comm. Math. Phys.}, {\bf 233}, 513--543, 2003.

\bibitem{ABS2009}
V.~E. Adler, A.~I. Bobenko, and Y.~B. Suris.
\newblock Discrete nonlinear hyperbolic equations. classification of integrable
  cases.
\newblock {\em Funct. Anal. Appl.}, {\bf 43}, 3--17, 2009.

\bibitem{BellonViallet1999} M. Bellon and C-M. Viallet, Algebraic Entropy, {\it Comm. Math. Phys.},
    {\bf 204}, 425--437, 1999.

\bibitem{Boll2011}
R.~Boll.
\newblock Classification of {3D} consistent quad-equations.
\newblock {\em J. Nonl. Math. Phys.}, {\bf 18}, 337--365, 2011.

\bibitem{Boll2012a}R.~Boll, Corrigendum Classification of 3D consistent quad--equations, {\it J. Nonl. Math. Phys.} {\bf 19}, 1292001 (3 pp), 2012.

\bibitem{Boll2012b}R.~Boll, Classification and Lagrangian structure of 3D consistent quad--equations, Ph. D. dissertation, 2012.

\bibitem{DemskoyViallet2012} D.K. Demskoy and C.-M. Viallet, Algebraic entropy for semi-discrete equations. 
    {\it J. Phys. A: Math. Theor.} {\bf 45},  352001 (10 pp), 2012.

\bibitem{gubbiotti_thesis} G. Gubbiotti Ph. D. Dissertation to be submitted in 2016.

\bibitem{GubScimLev2015} G. Gubbiotti, C. Scimiterna and D. Levi, On quad equations consistent on the cube, submitted to {\it J. Phys. A: Math. Theor.} .

\bibitem{Gallipoli15} G. Gubbiotti, C. Scimiterna and D. Levi, Linearizability and fake Lax pair for a consistent
around the cube nonlinear non--autonomous
quad--graph equation, submitted to {\it Teoreticheskaya i Matematicheskaya Fizika}.


\bibitem{Levi1981}D. Levi, Nonlinear differential-difference equations as B\"acklund transformations, {\it J. Phys. A: Math. Gen.}
{\bf 14}, 1083--1098,  1981.


\bibitem{LeviPetreraScimiterna2008}
D.~Levi, M.~Petrera, C.~Scimiterna, and R.~I. Yamilov.
\newblock On Miura transformations and Volterra-type equations associated with
  the {Adler}--{Bobenko}--{Suris} equations.
\newblock {\em SIGMA}, {\bf 4}, 077 (14pp), 2008.

\bibitem{LeviWinternitzYamilov}D. Levi, P. Winternitz and R.~I. Yamilov, Symmetries of the continuous and
discrete Krichever--Novikov equation, {\it  SIGMA}, {\bf 7}, 097 (21 pp),  2011.

\bibitem{LeviYamilov1995} D. Levi and R.~I. Yamilov, Classification Of Evolutionary Equations On The Lattice
I. The General Theory, arXiv:solv-int/9511006 v1 16 Nov 1995.

\bibitem{LeviYamilov1997}
D.~Levi and R.~I. Yamilov.
\newblock Conditions for the existence of higher symmetries of the evolutionary
  equations on the lattice.
\newblock {\em J. Math. Phys.}, {\bf 38}, 6648--6674, 1997.

\bibitem{Hydon2007}
O.~G. Rasin and P.~E. Hydon.
\newblock Symmetries of integrable difference equations on the quad-graph.
\newblock {\em Stud. Appl. Math.}, {\bf 119}, 253--269, 2007.



\bibitem{viallet2014} C. M. Viallet, Algebraic entropy for differential--delay equations \texttt{arXiv:1408.6161}, 2014.

\bibitem{Xenitidis2009}P. D. Xenitidis, Integrability and symmetries of difference equations: the Adler-Bobenko-Suris case, Proc. 4$^th$ Workshop Group Analysis of Differential Equations
and Integrable Systems, (2009) 226--242 (arXiv:0902.3954).

\bibitem{XenitidisPapageorgiou2009}
P.~D. Xenitidis and V.~G. Papageorgiou.
\newblock Symmetries and integrability of discrete equations defined on a
  black and white lattice.
\newblock {\em J. Phys. A: Math. Theor.}, {\bf 42}, 454025 (13 pp), 2009.

\bibitem{Yamilov1983} R.~I. Yamilov  \newblock Classification of discrete evolution equations \newblock {\em Uspekhi Mat. Nauk} {\bf 38}, 155--156, 1983 (in Russian).

\bibitem{Yamilov1984}R. I. Yamilov,  {\it Discrete equations of the form $\dot u_n = F(u_{n-1}, u_n, u_{n+1}) \; n \in Z$ with an infinite number
of local conservation laws} Dissertation for Candidate of Science (Ufa: Soviet Academy of Sciences) 1984 (in
Russian).

\bibitem{Yamilov2006}R.~I. Yamilov, Symmetries as integrability criteria for differential
difference equations, {\it J. Phys. A: Math. Gen.} {\bf 39},  R541--R623, 2006.
\end{thebibliography}

\end{document}